\documentclass[fleqn,twoside,dvips]{mystyle}
\usepackage[tbtags]{amsmath}
\usepackage{amssymb,bm}
\usepackage{graphicx}
\newcommand{\putfig}[2]{\includegraphics[width=#1\linewidth]{#2}}

\usepackage{txfonts}
\def\bm#1{\mbox{\boldmath $#1$}}

\usepackage{cite}

\newcommand{\pp}[2]{\frac{\partial #1}{\partial #2}}
\newcommand{\cE}{\mathcal{E}}

\def\br{{\bm{r}}}
\def\bp{{\bm{p}}}

\def\Td{T_d}
\newcommand{\etal}{\textit{et al}}

\def\<{\langle}
\def\>{\rangle}
\def\eps{\varepsilon}

\begin{document}

\title{Periodic orbit bifurcations and local \\
symmetry restorations in exotic-shape \\
nuclear mean fields}

\author{
{\large\textbf{\textsf{Ken-ichiro Arita}}}\\[1em]
{\rm Department of Physics, Nagoya Institute of Technology,
Nagoya 466-8555, Japan}\\[1em]
{\rm E-mail: arita@nitech.ac.jp}
}
\pacs{21.60.-n, 36.40.-c, 03.65.Sq, 05.45.Mt}

\date{Submitted: 28 February 2017 \\
Published: 20 June 2017 in Phys. Scr. {\bf 92}, 074005}

\iopabs{
The semiclassical origins of the enhancement of shell effects in exotic-shape
mean-field potentials are investigated by focusing attention on the
roles of the local symmetries associated with the periodic-orbit
bifurcations.
The deformed shell structures for four types of pure octupole shapes in
the nuclear mean-field model having a realistic radial dependence are
analyzed.  Remarkable shell effects are shown for a large $Y_{32}$
deformation having tetrahedral symmetry.  Much stronger shell effects
found in the shape parametrization smoothly connecting the sphere and
the tetrahedron are investigated from the view point of the classical-quantum
correspondence.  The local dynamical symmetries associated with the
bridge orbit bifurcations are shown to have significant roles in
emergence of the exotic deformed shell structures for certain combinations
of the surface diffuseness and the tetrahedral deformation parameters.
\\[5pt]
Keywords: periodic orbit theory, nuclear tetrahedral deformation,
shell structure
}

\maketitle

\section{Introduction}

The relative stabilities of atomic nuclei are essentially determined by
single-particle shell effects\cite{MayJen,BohrMot1,RingSchuck}.
The properties of nuclear shell
structures are strongly dependent on the symmetries of the mean-field
potential.  In the nuclear potentials with a spherical symmetry,
shell structures are greatly developed and one has
distinct magic numbers corresponding to shell closures
($N,Z=2,8,20,28,50,82,126,\cdots$).  As the origin of such remarkable
shell structures, both the spherical symmetry and the effect
of an approximate dynamical symmetry called pseudo SU(3) are
important\cite{Bahri}; the latter causes bunchings of levels
with different angular momenta.

The roles of dynamical symmetries are more important in establishing
deformed shell structures.  The origins of super- and hyperdeformations,
denoting extraordinarily large quadrupole deformations whose axis
ratios are 2:1 and 3:1, can be simply accounted for as the results of
the dynamical symmetries of the harmonic oscillator models with rational
frequency ratios\cite{Nazarewicz}.  In more realistic mean-field
potentials having sharper surfaces, approximate (weakly broken)
dynamical symmetries should play roles.
In general, the existence of such dynamical symmetries is
hidden and not found by just looking at the Hamiltonian.

We would like to point out the fact that the dynamical symmetries are
reflected in the stability properties of the classical periodic orbits
(POs).  Therefore, the semiclassical periodic-orbit theory (POT)
\cite{Gutzwiller,BaBlo,StrMag77,BBText} is a good tool to investigate
the relationship between the quantum shell effect and the dynamical
symmetries.  We will pay special attention to the bifurcations of the
classical POs, which are related to the local dynamical symmetries
restored in the phase space around the orbits\cite{Arita2016}.

Although the quadrupole deformations play the most important
role in lowering the energies of nuclei away from
the spherical shell closures, other
types of shape degrees of freedom, such as octupoles, also become
significant in certain particle-number regions\cite{Butler96}.  Since
the shell effects associated with octupole deformations are generally
weaker than those for quadrupole ones, protons and neutrons should
play constructive roles to realize such deformations.  Recent
development of experiments opens the door to exploration of various
combinations of $N$ and $Z$ considerably far from the stability line,
and the possibilities of tetrahedral-shape nuclei attract greater
attention from both the theoretical and experimental
sides\cite{Dudek2006,Dudek2007,Yamagami2001,Tagami2013}.  In several
theoretical mean-field calculations, it is indicated that the $Y_{32}$
shape degree of freedom having tetrahedral symmetry is very important
and the potentials with the tetrahedral-type deformations show strong
shell effects\cite{Hamamoto,Dudek2006,Reimann}.
In addition to the geometrical degeneracies due to the
point-group symmetry, strong bunching of levels belonging to
different irreps are found.  Hamamoto \etal\ tried to understand the
reason for this gross shell effect by looking at how the degeneracy of
$\varDelta l=3$ levels in the spherical limit resolves by the octupole
perturbations\cite{Hamamoto}, but its relation to dynamical symmetries
has not been considered.  It is very interesting to notice that, in a
Woods-Saxon (WS) type potential without spin-orbit coupling, the
tetrahedral magic numbers are identical to those of the spherical
harmonic oscillator (HO) having the SU(3) dynamical
symmetry\cite{Hamamoto,Reimann}.  This may suggest that SU(3)
dynamical symmetry is partially restored for certain combinations of
the sharpness of the potential surface and the tetrahedral-type
deformation.

In this paper, we first recapitulate the analysis of deformed shell
structures for four types of pure octupole shapes using a
nuclear mean-field
potential model with a realistic radial dependence.  We will show that
anomalously strong shell effects are found for a shape parametrization
interpolating the sphere and tetrahedron.  Using the semiclassical
POT, we will analyze the origin of the above
tetrahedral shell structures and point out the important role by a
special type of the PO bifurcation, called a bridge-orbit
bifurcation\cite{Arita2012}.
The relationship between the dynamical symmetry restoration
and PO bifurcations is discussed, and the semiclassical
origins of exotic tetrahedral shell structures are clarified.

\section{Shell structures in tetrahedral-shape potentials}

In this section, the radial power-law (RPL) potential model is
introduced to approximate the realistic WS type potential.  With this
simplified model, we examine the shell structures in octupole shapes
and show that the $Y_{32}$ type deformations having tetrahedral $\Td$
symmetry give rise to strong shell effects.

\subsection{The radial power-law (RPL) potential model}

\begin{figure}
\centering
\putfig{.75}{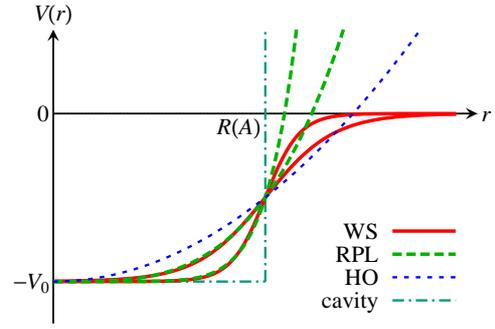} %wsfit.eps
\caption{\label{fig:wsfit}
Comparison of the Woods-Saxon (WS), radial power-law (RPL), harmonic
oscillator (HO, $\alpha=2$) and cavity (infinite well, $\alpha=\infty$)
potentials.}
\end{figure}

The nuclear mean-field potential is approximately described by the WS
model
\begin{equation}
V_{\rm ws}=-\frac{V_0}{1+\exp\frac{r-R(A)}{a}}
\end{equation}
with depth $V_0\approx 50\,\mbox{MeV}$, radius $R(A)\approx
1.3A^{1/3}\,\mbox{fm}$ for the mass number $A$, and surface
diffuseness $a\approx 0.7\,\mbox{fm}$.  This potential can be
approximated by a HO potential for light nuclei, while it is rather
closer to a square-well potential (which is further approximated by an
infinite well) for heavier nuclei.  In order to describe the nuclei in
a wide range of mass numbers, we propose the RPL potential\cite{Arita2012}
\begin{equation}
V(\br)=-V_0+U_0\cdot\left(\frac{r}{R_0}\right)^\alpha, \label{eq:valpha}
\end{equation}
which nicely approximates the inner part ($r\lesssim R(A)$) of the
WS potential (see figure~\ref{fig:wsfit}).
Eliminating the constant term in the potential (\ref{eq:valpha}), we
define our model Hamiltonian as
\begin{equation}
H=\frac{p^2}{2M}+U_0\cdot\left(\frac{r}{R_0f(\theta,\varphi)}\right)^{\alpha}
\label{eq:Hamil_RPL}
\end{equation}
with nucleon mass $M$, length unit $R_0$, energy unit
$U_0=\hbar^2/MR_0^2$, and a dimensionless function $f(\theta,\varphi)$
which describes the shape of the equi-potential surface.  The sharpness
of the potential surface is controlled by the power parameter
$\alpha$.  The advantage of taking the RPL model in place of WS is the
scaling property which makes our semiclassical analysis highly simple
(see section~\ref{sec:semiclassical} below).  The Hamiltonian can be
also extended to take account of the spin-orbit coupling keeping
some advantages of the scaling property\cite{Arita2016}, but we neglect
it in the current study for simplicity.

\subsection{Evolution of the shell structure with
pure octupole deformations}
\label{sec:shell_oct}

Hamamoto \etal\ have examined the deformed shell structures for the four
types of pure octupole shapes by adding the octupole field to the
modified oscillator potential without spin-orbit coupling as
\begin{equation}
V=\frac12m\omega_0^2r^2\left\{1+\eps_{3\mu}
\frac{Y_{3\mu}+Y^*_{3\mu}}{\sqrt{2(1+\delta_{\mu 0})}}\right\}+V_{ll},
\label{eq:v_hamamoto}
\end{equation}
and have obtained remarkable shell gaps for $Y_{32}$ deformation
around $\varepsilon_{32}\approx 0.5$~\cite{Hamamoto}.
In the central part of the potential (\ref{eq:v_hamamoto}),
the shape of the equi-potential surface reads
\begin{equation}
R(\theta,\varphi;\eps_{32})=R_0
\left[1+\eps_{32}\frac{Y_{32}+Y_{32}^*}{\sqrt{2}}\right]^{-1/2}.
\label{eq:shape_mho}
\end{equation}
Dudek \etal\ investigated the tetrahedral shell structures using
the realistic WS potential with spin-orbit
coupling\cite{Dudek2006,Dudek2007}.
The shape of the potential surface is parametrized as
\begin{equation}
R(\theta,\varphi;t_3)=R_0\Bigl\{1+t_3(Y_{32}+Y_{3-2}^*)+\cdots\Bigr\}.
\label{eq:shape_WS}
\end{equation}
They also obtained remarkable shell gaps at a large $Y_{32}$ deformation
$t_3\approx 0.3$ in the single-particle level diagram.  Thus, one sees
that the tetrahedral-type shape degrees of freedom play significant
roles in the nuclear dynamics.

\begin{figure}
\centering
\putfig{1}{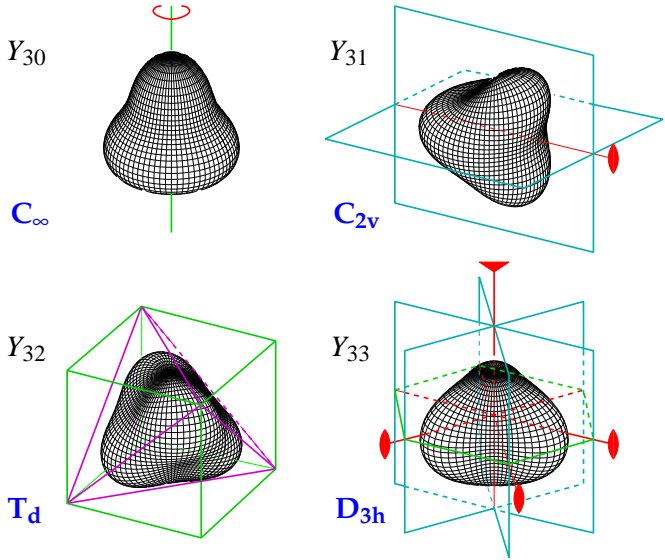} %voct_sym2.eps
\caption{\label{fig:y3shapes}
Shapes and symmetries of the potential surfaces for the
pure octupole deformations (\ref{eq:y3shapes}) with $\beta_{3\mu}=0.3$.}
\end{figure}

\begin{figure}[p]
\centering
\putfig{0.83}{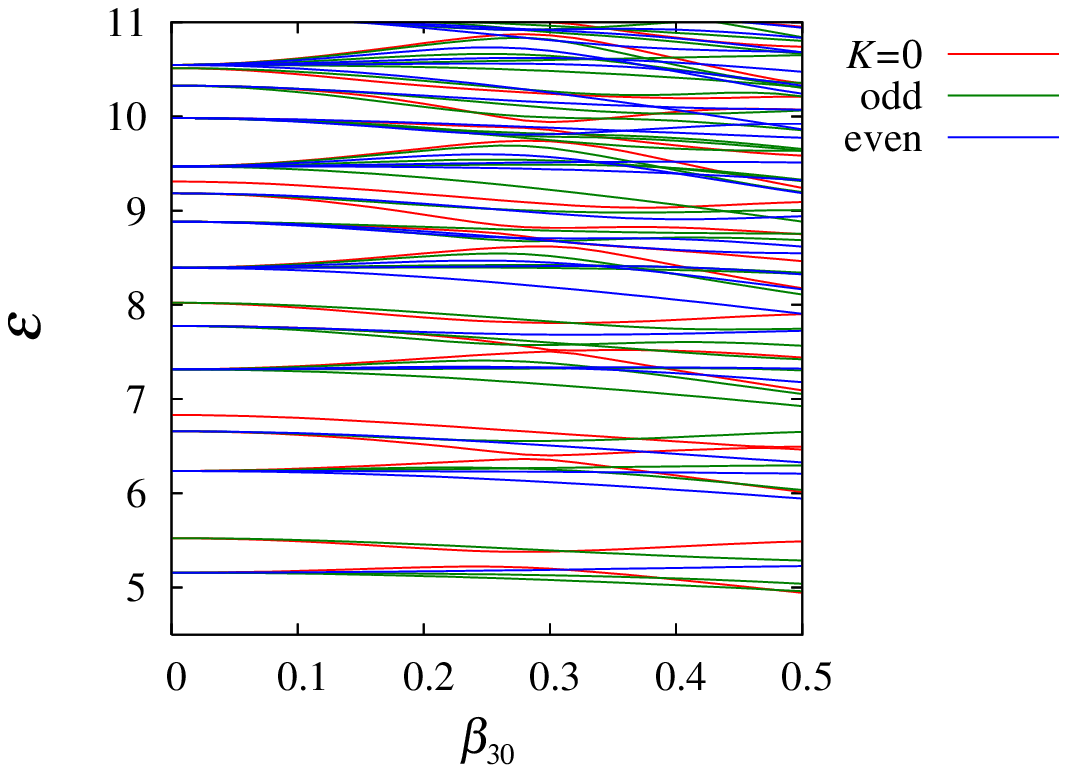} \\ %nils30_a050.eps
\putfig{0.83}{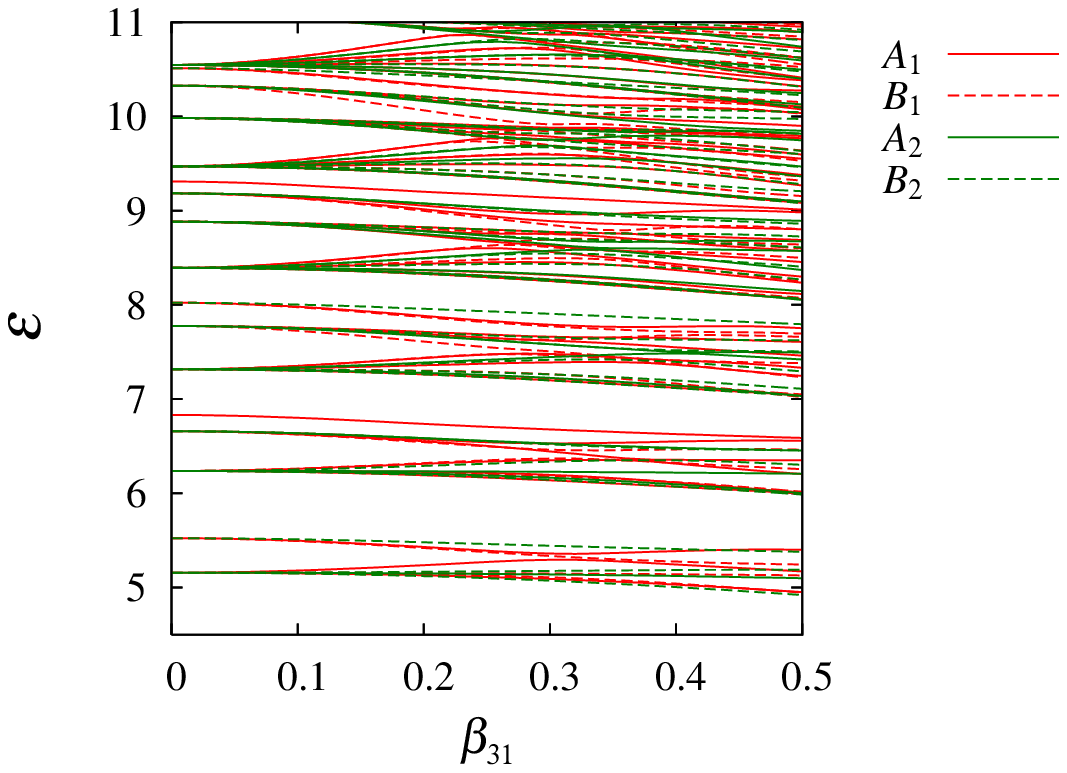} \\ %nils31_a050.eps
\putfig{0.83}{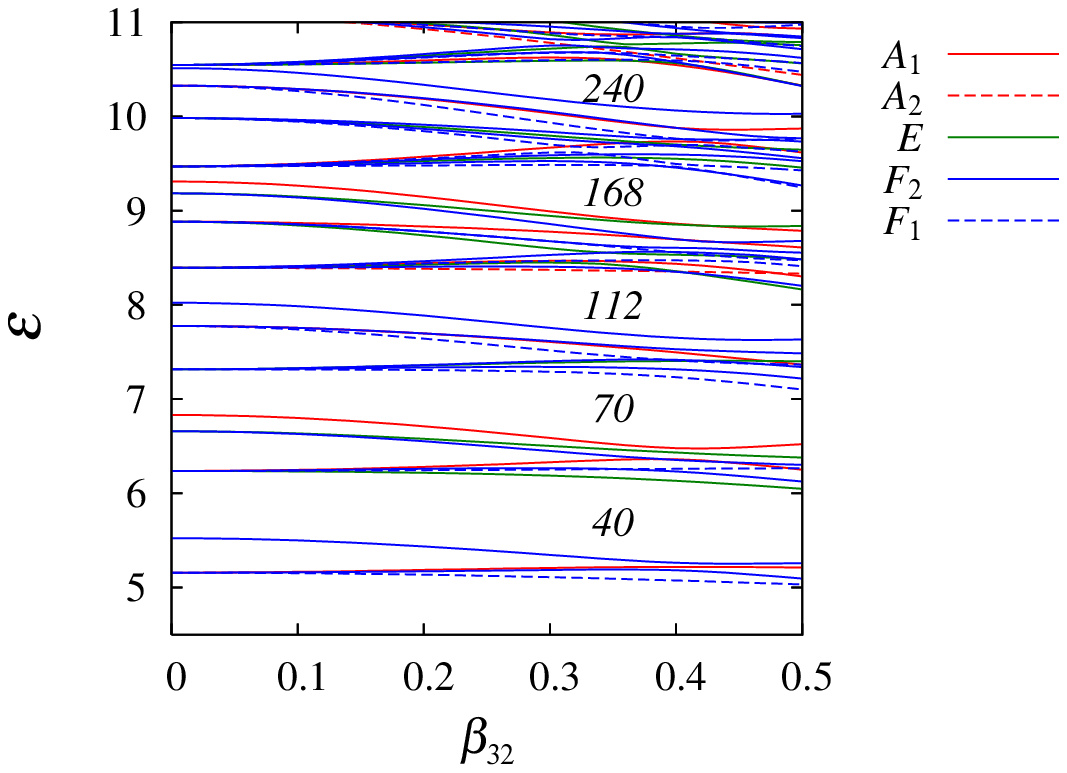} \\ %nils32_a050.eps
\putfig{0.83}{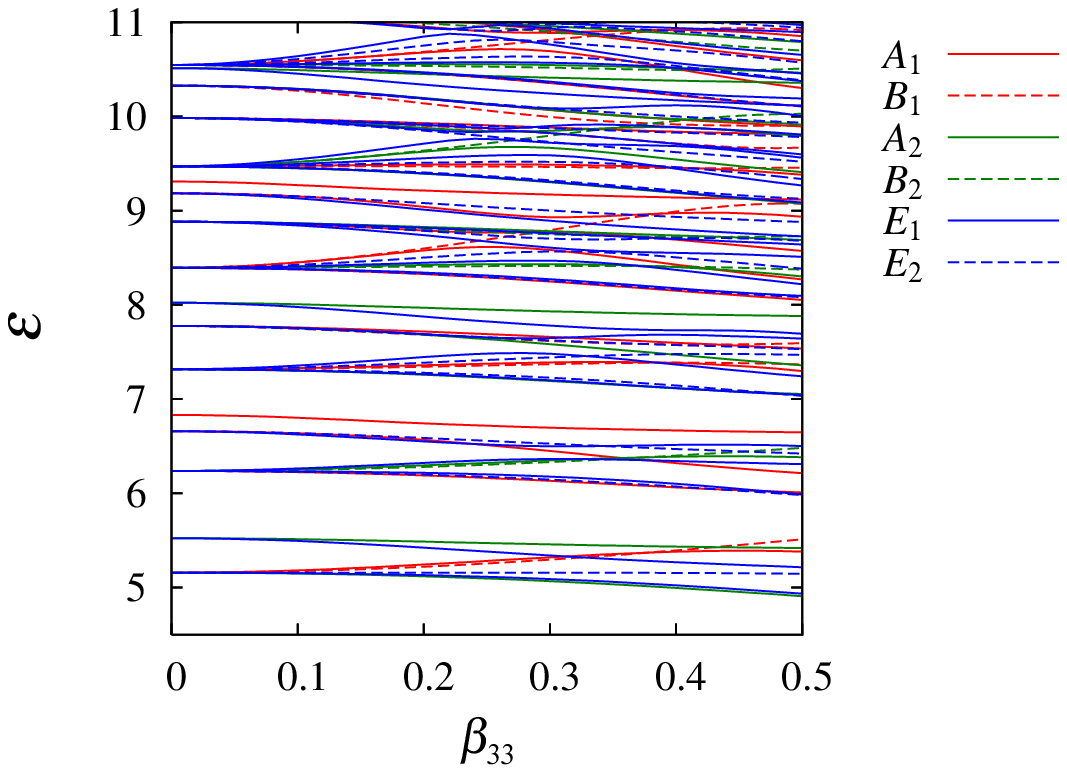} \\ %nils33_a050.eps
\caption{\label{fig:sps_oct}
Single-particle level diagrams of the RPL potential model with the
power parameter $\alpha=5.0$ for four types of pure octupole deformations.
Values of the scaled energy $\cE$ (see equation~(\ref{eq:scaledvar}))
for the single-particle eigenstates,
$\cE_i=(e_i/U_0)^{1/\alpha+1/2}$, are plotted as
functions of the octupole parameter $\beta_{3\mu}$.
For $\mu=0$, levels are classified by the magnetic quantum number
$K$.  For $\mu\ne 0$, levels are classified by the irreps of the
point group symmetries.  1-dimensional ($A_i$, $B_i$), 2-dimensional ($E_i$)
and 3-dimensional ($F_i$) irreps are singly, doubly and triply
degenerated, respectively.  In the 3rd panel from the top, the
particle number of the shell closures at large $\beta_{32}$ are
indicated with italics.}
\end{figure}

The shape parametrizations (\ref{eq:shape_mho}) and
(\ref{eq:shape_WS}) are equivalent for small $Y_{32}$ deformations but
become considerably different for large deformations.
These parametrizations can be generalized into the form
\begin{equation}
R(\theta,\varphi)=R_0\left[1+k\beta_{3\mu}
\frac{Y_{3\mu}+Y_{3\mu}^*}{\sqrt{2(1+\delta_{\mu 0})}}
\right]^{1/k},
\end{equation}
where (\ref{eq:shape_mho}) and (\ref{eq:shape_WS}) corresponds to the
case $k=-2$ and $k=1$, respectively, with
$\beta_{32}=2\eps_{32}=\sqrt{2}t_3$.  We determine the parameter $k$
to minimize the surface area under volume conservation condition,
and obtained the value $k\approx 0$ for all four types of
octupole deformations.  Using the relation $\lim_{k\to
0}(1+kx)^{1/k}=e^{x}$, we define the shape function $f$ in
equation~(\ref{eq:Hamil_RPL}) as
\begin{gather}
f(\varOmega;\beta_{3\mu})=\exp\left[\beta_{3\mu}
 \frac{Y_{3\mu}+Y_{3\mu}^*}{\sqrt{2(1+\delta_{\mu 0})}}
 \right]. \label{eq:y3shapes}
\end{gather}
The shapes of the equi-potential surfaces are displayed in
figure~\ref{fig:y3shapes}.
Each shape has its own continuous or discrete point-group symmetry and
the quantum levels are classified by the irreducible representations
(irreps) of the group.  The $Y_{30}$ shape is axially symmetric
($C_\infty$) and the single-particle levels are classified by the
magnetic quantum number $K$.  All the levels except $K=0$ are doubly
degenerate due to the time-reversal symmetry.  The $Y_{31}$ and $Y_{33}$
shapes have discrete symmetries $C_{2v}$ and $D_{3h}$, respectively.
The group $C_{2v}$ has only 1-dimensional irreps while $D_{3h}$ has
2-dimensional irreps in addition which provide doubly degenerate
levels.  The $Y_{32}$ shape has the tetrahedral symmetry $\Td$, which consists
of 24 symmetry transformations and has 3-dimensional irreps.  The
levels belonging to the 3-dimensional irreps are triply degenerate and,
thus, one can expect larger shell effects compared with the other three.
For each deformation, single-particle level diagrams are shown in
figure~\ref{fig:sps_oct}.
With increasing $\beta_{3\mu}$, one finds a remarkable enhancement of
the shell effect at a large $Y_{32}$ deformation around $\beta_{32}=0.3\sim
0.4$, where large and regular equi-distant shell gaps appear.  The
particle numbers corresponding to the shell closures are $2, 8, 20,
40, 70, 112, 168, 240, \cdots$, which are identical to those for the
spherical HO model.  This coincidence is hard to believe as just
incidental, and we expect a kind of a symmetry restoration taking place
for this octupole potential.

\subsection{Anomalous shell effect at large tetrahedral deformation}
\label{sec:shell_tetra}

\begin{figure}
\centering
\putfig{0.6}{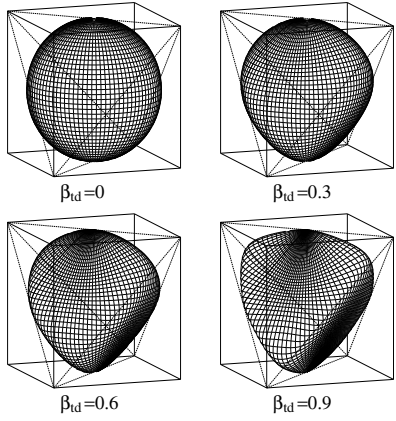} %vtetra.eps
\caption{\label{fig:vtetra}
Shapes of the equi-potential surfaces with the shape function
(\ref{eq:vtetra}) for several values of the tetrahedral parameter
$\beta_{\rm td}$.}
\end{figure}

\begin{figure}
\centering
\putfig{0.6}{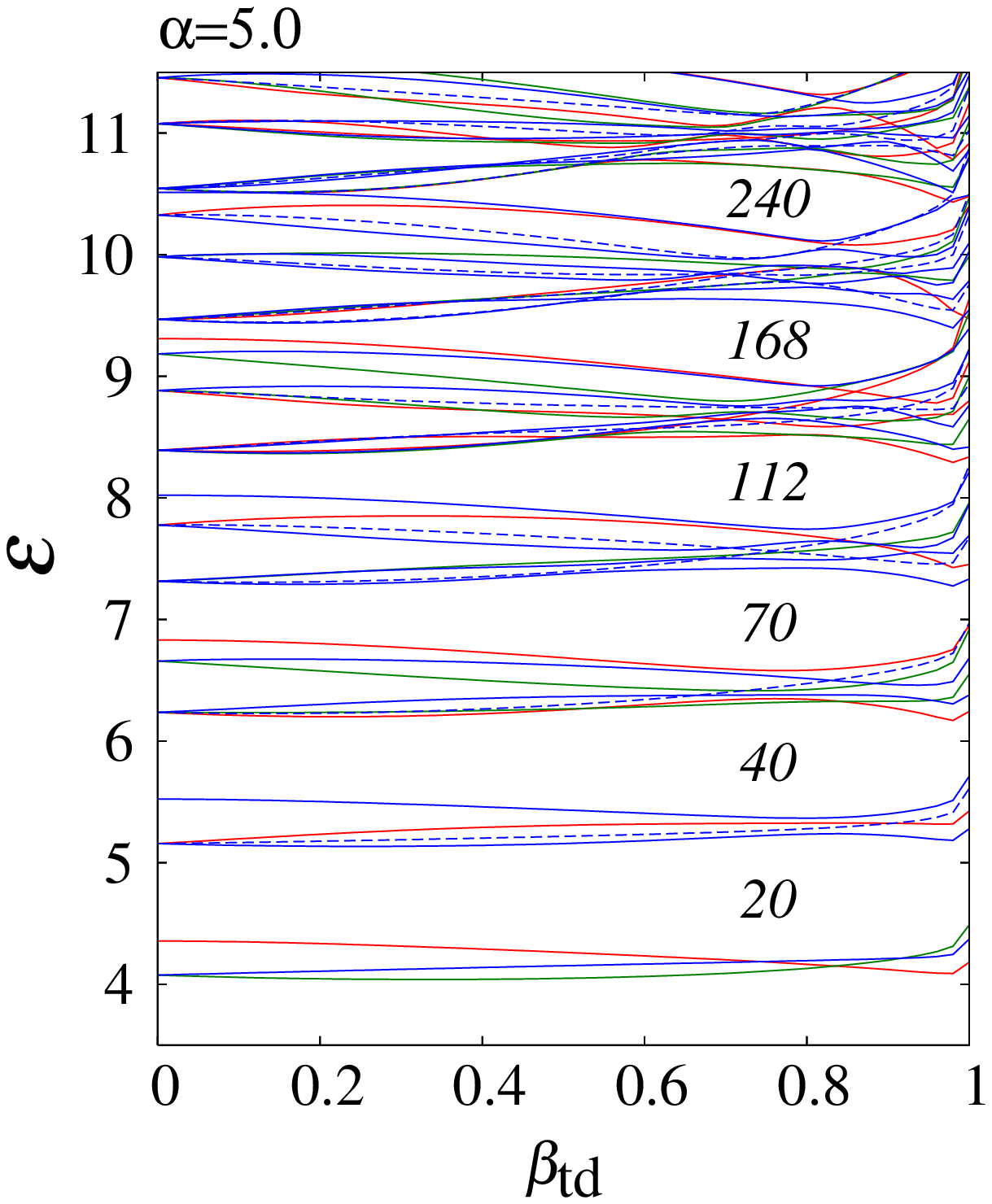} \\ %sps_tetra_a50r.eps
\putfig{0.6}{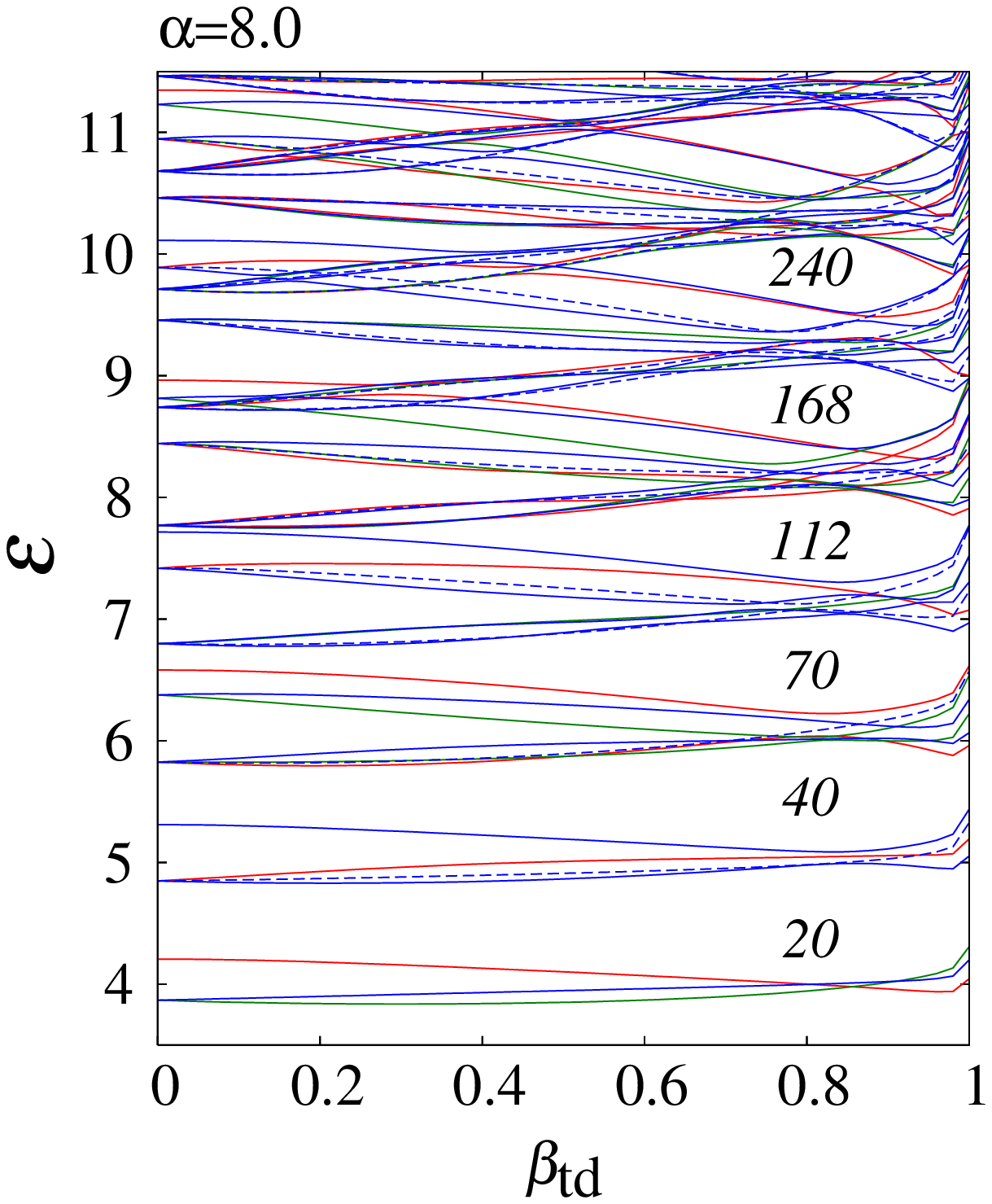}    %sps_tetra_a80r.eps
\caption{\label{fig:sps_tetra}
Single-particle level diagrams of the RPL potential models with radial
power parameter $\alpha=5.0$ (top panel) and $8.0$ (bottom panel)
for tetrahedral deformation.  Scaled energy levels are plotted as the
functions of tetrahedral parameter $\beta_{\rm td}$.
The levels are classified by the irreps of the $\Td$ group as 
for the $Y_{32}$ deformation in figure~\ref{fig:sps_oct}.
The particle numbers
corresponding to the shell closures at large $\beta_{\rm td}$ are
indicated with italics.}
\end{figure}

In the pure $Y_{32}$ deformation, potential surface suffers
concavity at large $\beta_{32}$ and the classical dynamics becomes
strongly chaotic.  In general, the quantum shell effect is moderate in
a classically chaotic system.  Thus, it might be favorable to evolve
a tetrahedral-type deformation keeping convexity of the surface in
obtaining a stronger tetrahedral shell effect.  We consider a shape
parametrization which smoothly interpolates the sphere and tetrahedron
with a single parameter $\beta_{\rm td}$\cite{AriMuku2014}.
The shape function $f$ is
obtained by the least positive root of the following quartic
equation at each pair of angles $(\theta,\varphi)$,
\begin{align}
f^2&+\beta_{\rm td}\left\{\frac12
 +\left(\frac{2}{15}P_{32}(\zeta)\sin 3\varphi\right)f^3 \right. \nonumber \\
&\left.-\left(\frac{1}{10}+\frac25P_{40}(\zeta)+\frac{1}{420}P_{44}(\zeta)
 \cos 4\varphi\right)f^4\right\}=1, \label{eq:vtetra}
\end{align}
with $\zeta=\cos\theta$.
$\beta_{\rm td}=0$ and 1 correspond to the sphere and tetrahedron,
respectively, and one can continuously change the shape from a sphere to
a tetrahedron
by varying $\beta_{\rm td}$ from 0 to 1, keeping the convexity
everywhere on the surface.  The shapes of the equi-potential surfaces
at several values of $\beta_{\rm td}$ are displayed in
figure~\ref{fig:vtetra}.
Single-particle levels of the RPL Hamiltonian with the tetrahedral
deformation are shown in figure~\ref{fig:sps_tetra}.
The tetrahedral shell effect in this shape parametrization is found to
be more pronounced than that of the pure $Y_{32}$ shape as we expected.
For sufficiently large values of the power parameter $\alpha$, we always
have the strong bunchings of levels at a large $\beta_{\rm td}$, whose
value becomes larger as $\alpha$ increases, and have the same
deformed magic numbers exactly identical to those of the spherical HO.
\begin{figure}
\centering
\putfig{.8}{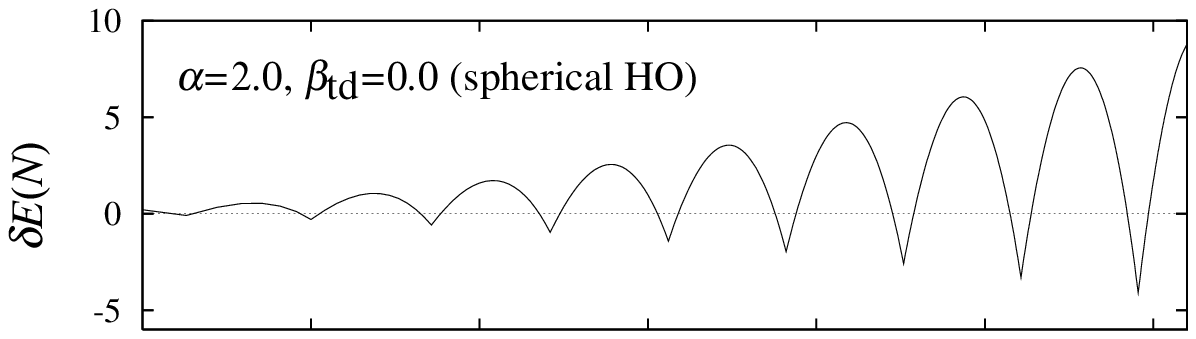} \\[-4mm] %sce_ho.eps
\putfig{.8}{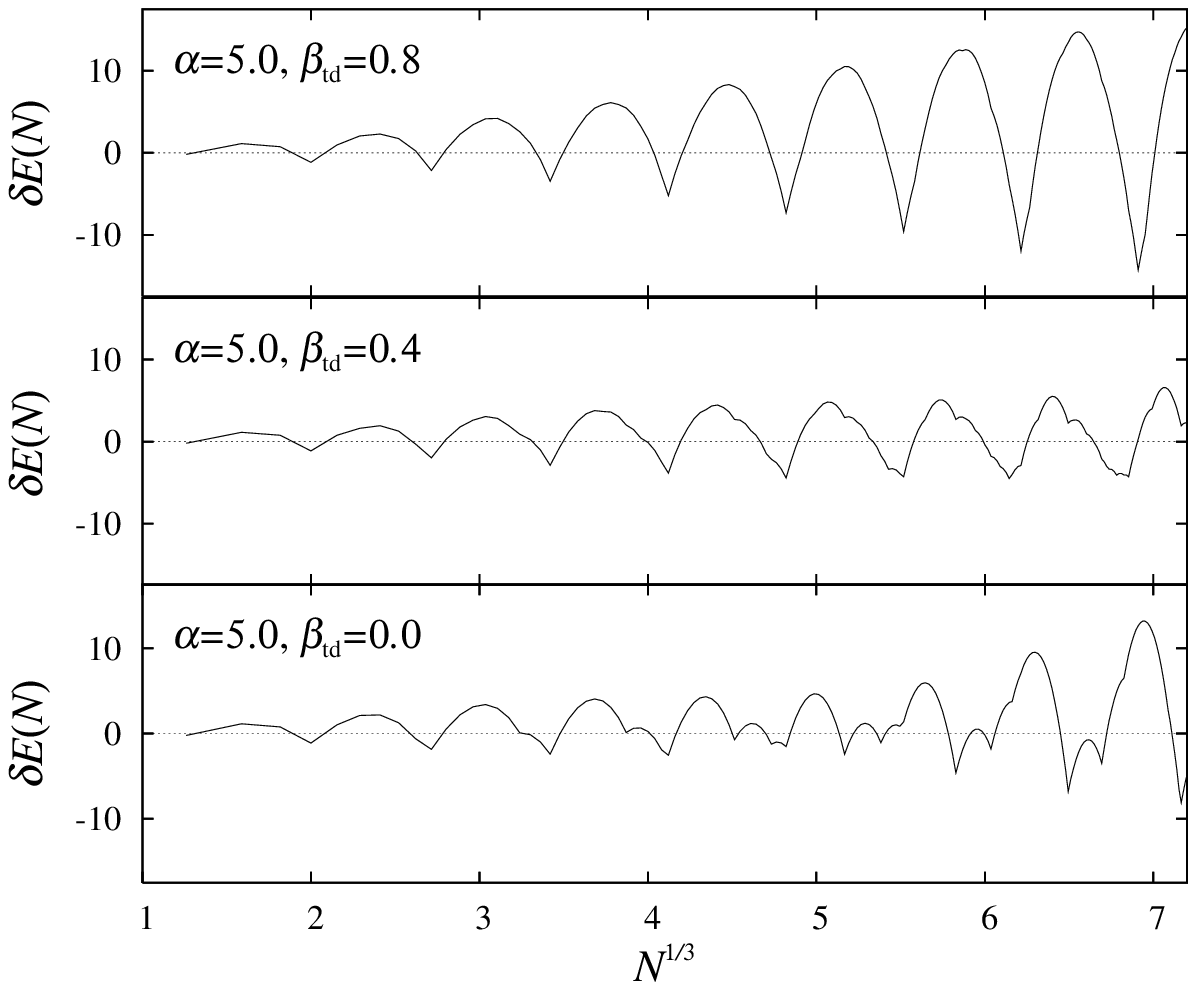}          %sce2_tetra_a50.eps
\caption{\label{fig:tetra_shell}
Shell energies $\delta E(N)$ for several values of the
tetrahedral parameters $\beta_{\rm td}$ in the RPL potential
model with $\alpha=5.0$ (lower 3 panels), and those for the
spherical HO (top panel, in a different energy unit).}
\end{figure}
In figure~\ref{fig:tetra_shell}, evolutions of the shell energies
$\delta E_{\rm sh}(N)$ are shown for several values of the tetrahedral
parameter $\beta_{\rm td}$.  A regular and strong shell effect is
found at a large tetrahedral deformation $\beta_{\rm td}\sim 0.8$, which
is quite similar to that for the spherical HO.

The above results suggest an emergence of a dynamical symmetry like
SU(3) restored by a certain combination of the sharp potential surface
and the tetrahedral deformation.  As we mentioned in the introduction,
dynamical symmetries are not easy to identify, as they are called hidden
symmetries, especially when they
are approximate ones, but one can find their signs in the properties of
classical POs.  In what follows, we will investigate the
semiclassical origin of the above outstanding shell effect at exotic
tetrahedral shapes using the POT.

\section{Semiclassical analysis of the tetrahedral shell structure}
\label{sec:semiclassical}

\subsection{Trace formula}

In the semiclassical trace formula\cite{Gutzwiller,BaBlo,StrMag77,BBText},
the single-particle level density is given by the sum over contributions
of the classical POs
\begin{equation}
g(e)=\bar{g}(e)+\sum_{\rm PO} A_{\rm PO}(e)\cos\left(
 \frac{1}{\hbar}S_{\rm PO}(e)-\frac{\pi}{2}\mu_{\rm PO}\right).
\label{eq:trace1}
\end{equation}
Here, $\bar{g}(e)$ is the average level density which is generally
a monotonous function of energy.  In the second term, representing the
oscillating part, the sum is taken over all the classical
POs including the repetitions of the primitive ones.
The amplitude factor $A_{\rm PO}$ is determined by the
period, degeneracy and stability of the orbit, $S_{\rm
PO}={\oint_{\rm PO}\bp\cdot d\br}$ is the action integral, and
$\mu_{\rm PO}$ is related to the Maslov index determined by
geometric property of the orbit.

The derivation of the trace formula is based on the path
integral representation of the transition amplitude
\begin{align}
K(\br'',t'';\br',t')&=\<\br''|e^{-i\hat{H}(t''-t')/\hbar}|\br'\> \nonumber \\
&=\int\mathcal{D}\br\exp\left[\frac{i}{\hbar}
\int_{t'}^{t''}L(\br,\dot{\br})dt\right],
\label{eq:Kpathint}
\end{align}
where $\hat{H}$ is the Hamiltonian operator,
$\mathcal{D}\br$ is the integration measure for the path $\br(t)$ with
$\br(t')=\br'$ and $\br(t'')=\br''$, and $L$ represents the Lagrangian
function.  In the semiclassical limit where action quantities are
sufficiently larger than the Plank's constant $\hbar$, the
path integral in (\ref{eq:Kpathint})
can be evaluated by the stationary-phase approximation (SPA):
for smooth functions $A(q)$ and $R(q)$ of a variable $q$, one has
\begin{equation}
\int A(q)e^{iR(q)/\hbar}dq \simeq
\sum_i\sqrt{\frac{2\pi i\hbar}{R''(q^*_i)}}A(q^*_i)e^{iR(q^*_i)/\hbar},
\label{eq:spa}
\end{equation}
where $q^*_i$ denotes the stationary point of the phase function $R(q)$
satisfying $R'(q^*_i)=0$.  The stationary condition for the path
integral (\ref{eq:Kpathint}) is nothing but the Hamilton principle
of the stationary action, and thus, the contributions associated with
the classical trajectories connecting $\br'$ and $\br''$ are
extracted.  Then, it is inserted into the level density
\begin{align}
g(e)&=\int \<\br|\delta(E-\hat{H})|\br\>d\br \nonumber \\
&=\frac{1}{2\pi\hbar}\int d\br\int_{-\infty}^\infty dte^{iet/\hbar}
K(\br,t;\br,0). \label{eq:trace_int}
\end{align}
Carrying out the integrations over $t$ and $\br(=\br'=\br'')$ in
Eq.~(\ref{eq:trace_int}) by the SPA, one obtains the
contribution of the POs at energy $e$.  It is important to
note that each term in the PO sum includes the contribution of
neighboring trajectories, which is reflected in the amplitude
factor $A_{\rm PO}$ as its dependency on the stability of the PO.

The energy dependence of the phase factor in Eq.~(\ref{eq:trace1})
is given by the action integral $S_{\rm PO}(e)$, which is a monotonically
increasing function of energy $e$, and thus, the contribution of
each PO gives a regularly
oscillating function of energy.  The period of the oscillations with
respect to energy,
\begin{equation}
\delta e=\frac{2\pi\hbar}{dS_{\rm PO}(e)/de}=\frac{2\pi\hbar}{T_{\rm PO}},
\end{equation}
is inversely proportional to the time period $T_{\rm PO}$
of the orbit, and
therefore, the gross structure of the level density (corresponding to large
$\delta e$) is associated with the short POs.

\begin{figure*}
\centering
\putfig{0.7}{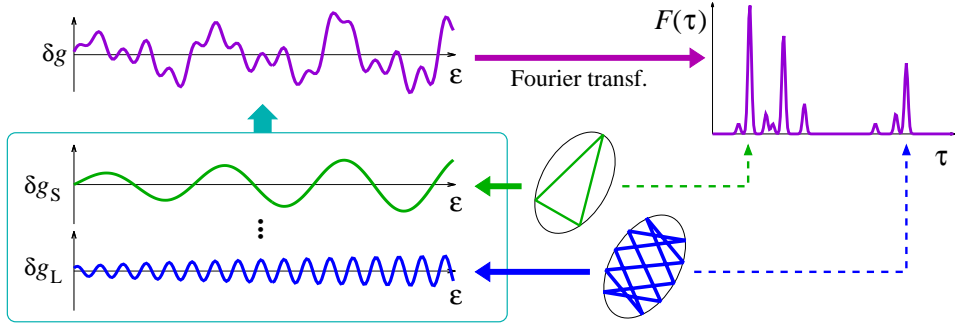} %potfig.ps
\caption{\label{fig:overview}
Illustration of the quantum-classical correspondence in the
single-particle level density via the Fourier transformation.
See the text for a detailed explanation.}
\end{figure*}

The outline of the semiclassical analyses of the shell structure
is illustrated in figure~\ref{fig:overview}.
On the left-hand side of the
figure, the top panel shows a typical example of the oscillating part
of the level density averaged over a certain resolution, which
generally shows a complicated pattern.
According to the semiclassical trace formula (\ref{eq:trace1}),
such an oscillation can be
always decomposed into several regular oscillations (as in the left
lower panels):
the slow oscillations as shown in $\delta g_{\rm S}$ are given by the
short orbits (having small periods $\tau$) which contribute to the
gross shell structures, while the rapid oscillations
as shown in $\delta g_{\rm L}$ are given by the longer
orbit (having large $\tau$) which contribute to the finer
shell structures.
As will be described below, the Fourier transform of the level
density (the right-hand panel) exhibits peaks at the periods of those
POs, and the heights of the peaks represent the magnitude
of the orbits' contributions to the level density.

Using the formula (\ref{eq:trace1}), one can derive the trace
formula for the shell energy\cite{Str75,StrMag76,BBText}
\begin{align}
\delta E(N)=
&\sum_{\rm PO}\left(\frac{\hbar}{T_{\rm PO}(e_F)}
 \right)^2 A_{{\rm PO}}(e_F) \nonumber \\
&\times \cos\left(\frac{1}{\hbar}S_{\rm PO}(e_F)
-\frac{\pi}{2}\mu_{\rm PO}\right),
\label{eq:trace_energy}
\end{align}
where $e_F$ is the Fermi energy determined as the function of the
particle number $N$ to satisfy
\begin{equation}
2\int_0^{e_F}g(e)de=N,
\end{equation}
taking account of the spin degeneracy factor 2.
One should note that, due to the extra factor proportional to
$(T_{\rm PO})^{-2}$ in (\ref{eq:trace_energy}),
contribution of long orbits are suppressed, and only
some shortest POs contribute to the shell energy.

It is known that the shell structure is generally sensitive to
the shape of the mean-field potential.  This can be understood from the
sensitivity of the stabilities of classical POs to the potential shape.
In particular, as will be discussed in the following
subsections, bifurcations of the POs
have strong effects on the oscillating part of the level density
at which the orbits change from stable to unstable.
Thus, we focus on the bifurcations of the short POs, which
play important roles in evolutions of the deformed shell structures.

\subsection{Classical periodic orbits in the RPL potential}
In order to clarify the semiclassical origin of the tetrahedral shell
structures, we first consider the properties of the classical POs
in the RPL potential.

Since the Hamiltonian (\ref{eq:Hamil_RPL})
is homogeneous both in momenta and in coordinates,
the following scaling relation holds:
\begin{equation}
c^{-1}H(c^{1/2}\bp,c^{1/\alpha}\br)=H(\bp,\br),
\end{equation}
and one can easily show the invariance of the equations of motion
under the scaling transformation
\begin{equation}
\bp\to c^{1/2}\bp, \quad
\br\to c^{1/\alpha}\br, \quad
t\to c^{1/\alpha-1/2}t,
\end{equation}
as energy $e\to ce$.
Therefore, one will find the same set of classical POs independent of
energy.  This scaling property highly simplifies the semiclassical
studies since the information on the POs at any energy $e$
can be obtained by those calculated at a certain energy, e.g., $e=U_0$.
\begin{figure}
\centering
\putfig{.9}{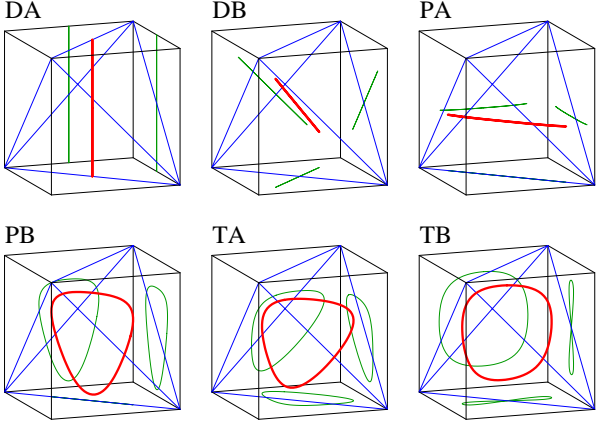} %po_tetra_a50.eps
\caption{\label{fig:po_tetra}
Some shortest POs in the tetrahedral RPL potential with
$\alpha=5.0$ and $\beta_{\rm td}=0.3$.
Their projections onto $(x,y)$, $(y,z)$ and $(z,x)$ planes are also
shown.  The tetrahedron represents the symmetry of the potential.}
\end{figure}
When the tetrahedral deformation is added to the spherical RPL
potential, the
diameter and circle orbits bifurcate into three branches each.  Two kinds
of the straight-line orbits DA and DB along the three $S_4$ axes and the
four $C_3$ axes, respectively, and curved self-retracing orbits PA in
each of the mirror planes emerge from the diameter family.  On
the other hand, two kinds
of three-dimensional rotational orbits TA, TB and planar ones PB
emerge from the circle family.  These six orbits are
displayed in figure~\ref{fig:po_tetra}.

The contributions of these POs to the shell effect
are manifested in the Fourier spectra of the quantum level density.
Using the scaling relation, the action integral is found to be
proportional to a simple power of energy:
\begin{equation}
S_{\rm PO}(e)=\left(\frac{e}{U_0}\right)^{\frac12+\frac{1}{\alpha}}
S_{\rm PO}(U_0)\equiv \cE\hbar\tau_{\rm PO}
\label{eq:scaledvar}
\end{equation}
with the energy unit $U_0=\hbar^2/MR_0^2$.  In the last equation, we
define the dimensionless ``scaled energy'' $\cE\equiv
(e/U_0)^{1/2+1/\alpha}$ and the energy-independent ``scaled period''
$\tau_{\rm PO}\equiv S_{\rm PO}(U_0)/\hbar$.  The trace formula
(\ref{eq:trace1}) is then rewritten in terms of these scaled
variables $(\cE,\tau)$ as
\begin{align}
g(\cE)&=g(e)\frac{de}{d\cE} \nonumber \\
&=g_0(\cE)+\sum_{\rm PO} A_{\rm PO}(\cE)
\cos(\tau_{\rm PO}\cE-\tfrac{\pi}{2}\mu_{\rm PO}) \label{eq:trace2}
\end{align}
Now we consider the Fourier transform of the scaled-energy level density
\begin{equation}
F(\tau)=\int d\cE\,e^{i\tau\cE}g(\cE).
\label{eq:fourier}
\end{equation}
For the quantum level density $g(\cE)=\sum_j\delta(\cE-\cE_j)$,
one has
\begin{equation}
F^{\rm qm}(\tau)=\sum_j e^{i\tau\cE_j}, \quad
 \cE_j=(e_j/U_0)^{\frac12+\frac{1}{\alpha}},
\label{eq:ftl_qm}
\end{equation}
which can be easily evaluated using the single-particle energy
spectra $\{e_j\}$ calculated quantum-mechanically.
Practically, we apply a triangular cut-off in which we multiply the
integrand by the function $(1-\cE/\cE_c)$ and evaluated the
integral over $0<\cE<\cE_c$.
Inserting the semiclassical trace formula (\ref{eq:trace2}), one
obtains the expression
\begin{equation}
F^{\rm cl}(\tau)=F_0(\tau)+\pi\sum_{\rm PO}
 e^{\frac{i\pi}{2}\mu_{\rm PO}}\widetilde{A}_{\rm PO}
 \delta(\tau-\tau_{\rm PO}).
\end{equation}
$F(\tau)$ will thus have successive peaks at the scaled periods
$\tau=\tau_{\rm PO}$ of the classical POs with the heights
proportional to the amplitude $A_{\rm PO}$ of the corresponding orbits.

\begin{figure}
\centering
\putfig{.9}{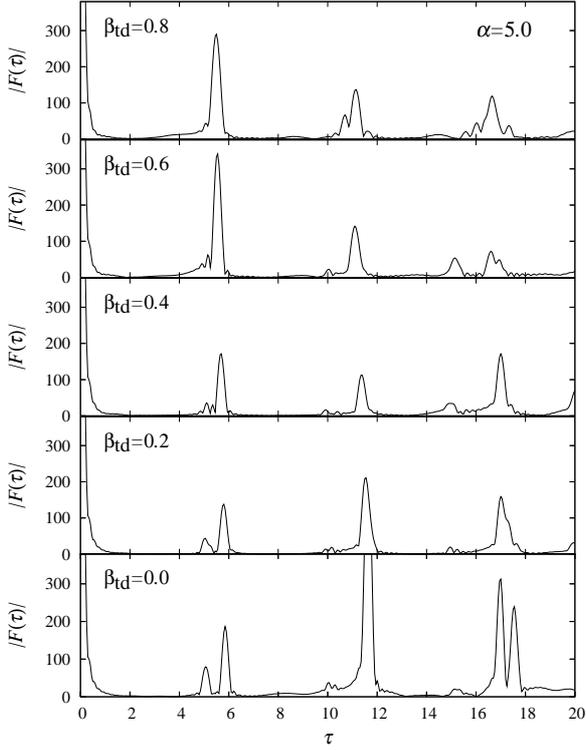} %ftl_tetra_a50.ps
\caption{\label{fig:ftetra}
Fourier spectra of the single-particle level densities in the RPL potential
($\alpha=5.0$).  Moduli of the Fourier transform (\ref{eq:ftl_qm}) are
plotted as functions of $\tau$ for several values of the tetrahedral
parameter $\beta_{\rm td}$.}
\end{figure}
Figure~\ref{fig:ftetra} shows the moduli of the Fourier transforms
$|F^{\rm qm}(\tau;\beta_{\rm td})|$ of the quantum
level density for the RPL model with $\alpha=5.0$ as functions of
$\tau$.  At the spherical shape, $\beta_{\rm td}=0$,
one sees peaks at the diameter ($\tau=5.06$) and circle ($\tau=5.84$) orbits.
With increasing tetrahedral deformation, the positions of
these two peaks approach each other and merge into
a single peak around $\tau=5.6$, which shows the
significant enhancement at $\beta_{\rm td}\simeq 0.6$.  As we show
below, this enhancement originates from the dynamical symmetry
restoration associated with a special type of the PO bifurcation.

\subsection{Bifurcation enhancement of the shell effect}

\begin{figure}
\centering
\putfig{0.6}{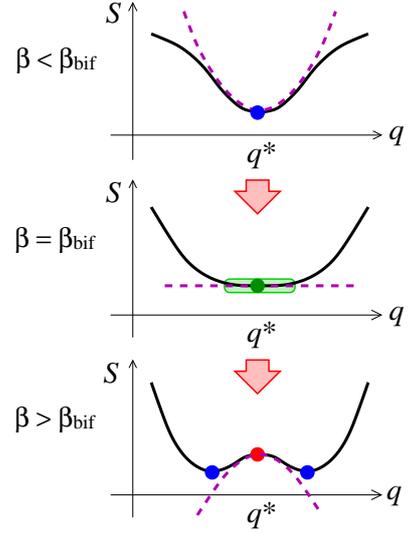} %bif2.eps
\caption{\label{fig:sbif}
Illustration of the PO bifurcation and emergence of a local
dynamical symmetry.  Dots denote POs corresponding to the
stationary point $q^*$ of the action $S(q)$.
Broken lines represent the quadratic approximation of $S(q)$ at the
stationary points $q^*$.
At the bifurcation
point $\beta=\beta_{\rm bif}$, family of quasi-periodic orbits (shaded
area) emerges around the periodic orbit $q^*$.}
\end{figure}

As mentioned above, shell energies are essentially
determined by the contributions of some shortest POs.
The amplitude $A_{\rm PO}$ depends on the shape mainly through the
stability factor as
\begin{equation}
A_{\rm PO}\propto\frac{1}{\sqrt{|\det(I-M_{\rm PO})|}},
\label{eq:amp_Gutz}
\end{equation}
where $M_{\rm PO}$ is the monodromy matrix which represents the
linear stability of the orbit as described below.
In the equi-energy surface $\{(\bp,\br)|H(\bp,\br)=e\}$, consider a
certain phase-space plane $\varSigma$ perpendicular to the orbit.  A classical
trajectory starting off the point $Z$ on $\varSigma$ will subsequently
cross $\varSigma$ again at $Z'$, and it defines a map
$\mathcal{M}$ called the Poincar\'{e} map, $Z'=\mathcal{M}(Z)$.  Periodic
orbits $Z_{\rm PO}$ are the fixed points of the map $\mathcal{M}$, and the
monodromy matrix is the linear part of $\mathcal{M}$ around the PO:
\begin{equation}
\mathcal{M}(Z_{\rm PO}+\delta Z)=Z_{\rm PO}+M_{\rm PO}\delta Z
+O(\delta Z^2).
\end{equation}
The factor $\det(I-M_{\rm PO})$ in (\ref{eq:amp_Gutz}) derives from the
trace integral carried out by the SPA, and it is
proportional to the curvature of the action $S(\br;e)$ along the
closed trajectory which starts from the point $\br$ with energy $e$
and returns to $\br$ again,
\begin{gather}
S(\br;e)=S(\br'',\br';e)|_{\br'=\br''=\br},\quad
S(\br'',\br';e)=\int_{\br'}^{\br''}\bp\cdot d\br. \nonumber \\
\end{gather}
The stationary points of $S(\br;e)$ correspond
to the POs since the final momentum coincides with the
initial momentum there:
\begin{align}
\pp{S(\br;e)}{\br}&=\left(\pp{S(\br'',\br';e)}{\br''}
+\pp{S(\br'',\br';e)}{\br'}\right)_{\br''=\br'=\br} \nonumber \\
&=\bp''-\bp'=0.
\end{align}
In general, the number of the stationary points changes when the
curvature $\det(\partial^2S/\partial\br\partial\br)$ changes its sign.
Therefore, the zeros of the curvature are
accompanied by the PO bifurcations.  Since the
stability factor $\det(I-M_{\rm PO})$ is proportional to the curvature,
the monodromy matrix $M_{\rm PO}$ has a unit eigenvalue at the
bifurcation point (or a pair of unit eigenvalues depending on the
bifurcation types), and the corresponding eigenvector $\delta
Z_1$ gives the new bifurcated PO since it satisfies the
periodic condition
\begin{equation}
\mathcal{M}(Z_{\rm PO}+\delta Z_1)\simeq Z_{\rm PO}+M_{\rm PO}\delta Z_1
=Z_{\rm PO}+\delta Z_1.
\end{equation}
A typical bifurcation scenario known as a ``pitchfork bifurcation'' is
illustrated in figure~\ref{fig:sbif}.
In this case, the number of the POs changes from 1 to 3.
Since the denominator in (\ref{eq:amp_Gutz})
approaches zero at the bifurcation point,
one expects an enhancement of the amplitude there.
The divergence of the amplitude at the bifurcation is due to the
break-down of the standard SPA (\ref{eq:spa}), and can be remedied by
the use of the uniform approximation\cite{SieberUA} which takes into
account the higher-order expansions of the phase function, or by the improved
SPA\cite{MagnerISPM} which keeps the finite integration limits.
This enhancement can be understood as the result of the local
dynamical symmetry associated with the bifurcation.
At the bifurcation point, the action function
is approximately flat around the stationary point in the certain direction.
This local invariance of the action against the change of coordinate
indicates an appearance of a local dynamical symmetry.  It generates
a locally degenerate family of quasi-periodic orbits, which will
make a coherent contribution to the trace integral.  Actually, it often
brings about a considerable
enhancement of the amplitude factor $A_{\rm PO}$.  This is the
bifurcation enhancement of the quantum shell effect, which we consider
as a significant semiclassical mechanism for exotic-shape states to be realized
in finite fermion systems.

\subsection{Bridge orbit bifurcation and local dynamical symmetry}

In the upper panel of figure 11, we plot the scaled periods $\tau_{\rm
PO}$ of some POs as functions of the tetrahedral
deformation parameter $\beta_{\rm td}$.
With increasing $\beta_{\rm td}$, the scaled periods $\tau$ of the
orbits DA and TB approach to each other.  At $\beta_{\rm td}=0.502$,
bifurcation of the diameter orbit DA takes place and a new
three-dimensional orbit TD emerges.  Subsequently, TD submerges into
the three-dimensional rotational orbit TB at
$\beta_{\rm td}=0.574$, shortly after its emergence.
\begin{figure}
\centering
\putfig{0.9}{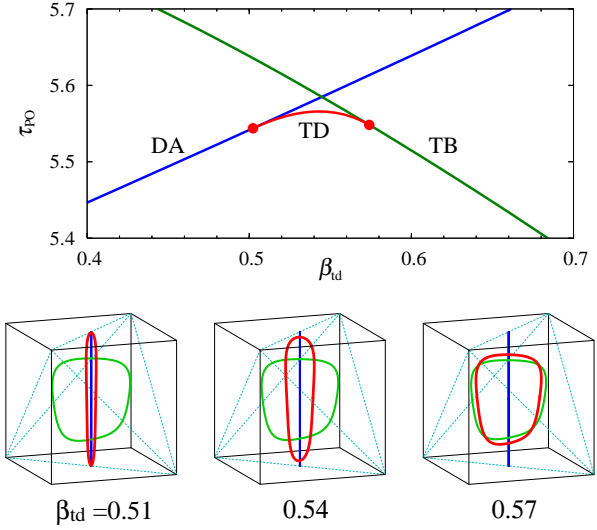} %bif_bridge2b.eps
\caption{\label{fig:bridge_tetra1}
Classical POs participating one of the bridge
bifurcations.  The bridge orbit TD
(red line) emerges from the diameter orbit DA (blue line)
at $\beta_{\rm td}=0.502$ and submerges into the three-dimensional
rotational orbit TB
(green line) at $\beta_{\rm td}=0.574$.  In the top panel,
scaled period $\tau_{\rm PO}$ of these orbits are plotted as functions
of $\beta_{\rm td}$.  In the lower panel, shapes of those orbits at
three values of $\beta_{\rm td}$ between the left and right ends
of the bridge are displayed.}
\end{figure}
This is what we call a bridge orbit bifurcation\cite{AriBra2008b}.  Note
that diametric DA and rotational TB are the orbits with minimum and
maximum angular momenta.  These greatly different orbits are connected by
the bridge orbit TD within a small change of the shape parameter
$\beta_{\rm td}$.  The same kind of the bridge
bifurcations but in different
pairs of POs take place almost simultaneously around $\beta_{\rm
td}=0.5\sim 0.6$, where we found a significant enhancement of the
shell effect.

\begin{figure}
\centering
\putfig{0.9}{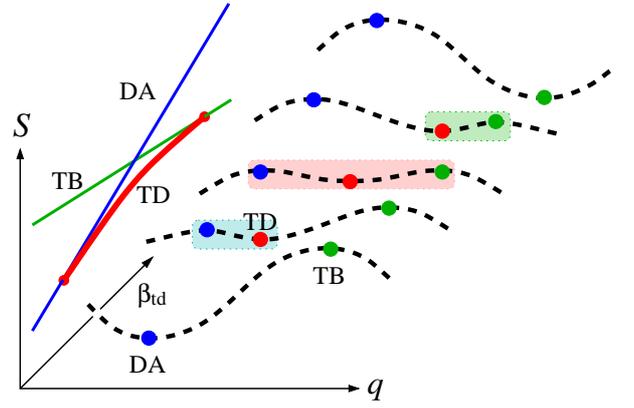} %bif_bridge2a.eps
\caption{\label{fig:bridge_tetra2}
Illustration of the bridge-orbit bifurcation scenario.  A family of
quasi-periodic orbits (indicated by the shaded area) is generated in
a large portion of the phase space connecting two POs,
DA and TB, which are distant from each other in the phase space.}
\end{figure}
Here, we would like to remark that
the bridge orbit bifurcation indicates a restoration of the dynamical
symmetry around the trail of the bridge orbit connecting two mutually distant
POs.  Figure~\ref{fig:bridge_tetra2} illustrate how such
a dynamical symmetry appears associated with the bridge orbit bifurcations.
A local quasi-periodic family is formed around the first bifurcation
where the DA bifurcates and the bridge TD emerges.
Such a local family is
also formed around the the second bifurcation where the orbit
TD submerges into
TB.  Although the stationary points corresponding to the periodic
orbits DA and TB are distinctly distant from each other, the two
bifurcation deformations at the ends of the bridge are close together,
which indicates
a restoration of the dynamical symmetry in a large portion of the
phase space including those three POs.  In the tetrahedral RPL model,
several bridge bifurcations take place simultaneously around
$\beta_{\rm td}=0.5\sim 0.6$, and the associated dynamical symmetry
will also show up around all their replicas
related by the 24 symmetry transformations
of the tetrahedral group $\Td$.
\begin{figure}
\centering
\putfig{1}{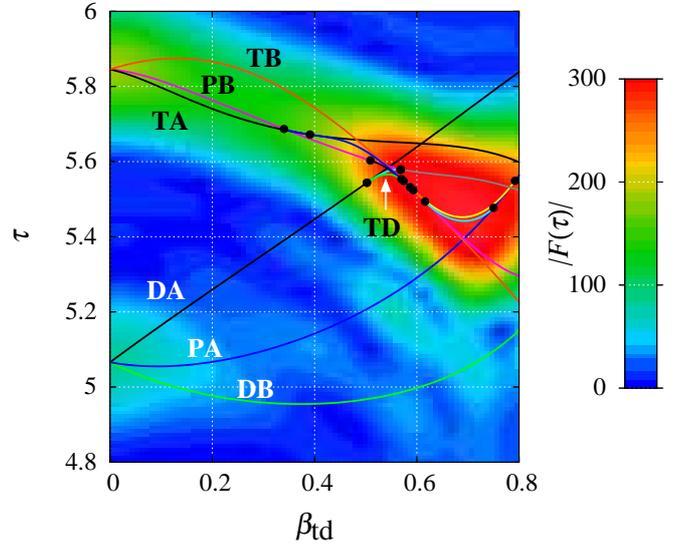} %fmap_tetra2.eps
\caption{\label{fig:fmap_tetra}
Quantum-classical correspondence in the Fourier transform of the
level density.  Color indicates the quantum Fourier amplitude
$|F^{\rm qm}(\tau;\beta_{\rm td})|$, and the
curves represent the scaled periods $\tau_{\rm PO}(\beta_{\rm td})$ of
the classical POs.}
\end{figure}
Figure~\ref{fig:fmap_tetra} shows the color map of the Fourier
amplitude $|F^{\rm qm}(\tau;\beta_{\rm td})|$ as functions of $\tau$
and $\beta_{\rm td}$.  The scaled periods $\tau_{\rm PO}$ of the
classical POs are also drawn.  The bifurcation points of
those POs are indicated by the solid circles.  This plot clearly shows
us that the bridge orbit bifurcations
around $\beta_{\rm td}=0.5\sim 0.6$ are the origin of the anomalously
strong shell effects found in Sec.~\ref{sec:shell_tetra}.
Considering together the agreement of
the deformed magic numbers to those of the spherical HO model, one may
expect a restoration of a large dynamical symmetry like SU(3).

\section{Summary}

The octupole deformed shell structures are investigated using the RPL
potential model.  Among the four types of pure octupole deformations,
particularly strong shell effects are found for certain combinations
of the power parameter $\alpha$ and the $Y_{32}$ deformation parameter
$\beta_{32}$.  The shell closures occur at particle numbers identical
to the magic numbers
in the spherical harmonic oscillator.  The shell effect is found
to be much more enhanced by taking the shape parametrization
which smoothly connects the sphere and tetrahedron.  Semiclassical
analysis of the shell structure revealed its origin as the dynamical
symmetry restoration associated with the bifurcations of bridge orbits
between several pairs of short POs taking
place at almost the same deformation $\beta_{\rm td}$.  Fourier
analysis of the level density clarified the correspondence between
those bridge orbit bifurcations and enhancement of quantum shell
effects.

In this peculiar type of the bifurcation scenario, two POs
which are locating apart from each other in the phase space are
connected by the bridge orbits, and families of quasi-periodic orbits
are formed around them.  In this sense, the bridge orbits play the
roles of generators for the relevant hidden symmetries.
If the six POs displayed in figure~\ref{fig:po_tetra} are considered as the
independent modes, the bridge orbit connecting two of those POs
generates a symmetry between the corresponding two modes, and the
number of the bridges may reflect the dimension of the symmetry.
We have at least seven bridges between the above six POs, appearing
almost simultaneously around $\beta_{\rm td}=0.5\sim 0.6$, and this
number is close to 8, namely, the dimension of SU(3).

The shape parametrization (\ref{eq:vtetra}) might be not realistic for
nuclei due to the large curvatures around the four vertices of the
tetrahedron.  However, the shell effect is so large and we expect
it to survive in a more
realistic parametrization of the nuclear shapes, and probably,
even after a reasonable spin-orbit coupling is switched on.
Semiclassical studies on effects of the spin-orbit coupling to the
tetrahedral shell structures are in progress.

\ack

The author thanks Prof. Jerzy Dudek and Prof. Kenichi Matsuyanagi for
valuable discussions and comments.

\end{document}